\documentclass[prb,preprint]{revtex4}
\usepackage{graphicx}
\makeatletter
\usepackage{epsfig}
\def\@dotsep{4.5}
\usepackage{dcolumn}
\usepackage{amsmath}
\makeatother

\usepackage{graphicx}
\usepackage{bm}
\usepackage{amsfonts}
\usepackage{amsbsy}
\usepackage{amssymb}
\usepackage[mathscr]{eucal}

\newcommand{\beq}{\begin{equation}}
\newcommand{\eeq}{\end{equation}}
\newcommand{\ba}{\begin{array}}
\newcommand{\ea}{\end{array}}
\newcommand{\bea}{\begin{eqnarray}}
\newcommand{\eea}{\end{eqnarray}}
\newcommand{\bseq}{\begin{subequations}}
\newcommand{\eseq}{\end{subequations}}

\begin{document}

\title{Temperature Dependence in Rainbow Scattering of Hyperthermal Ar Atoms from LiF(001)}

\author{W. W. Hayes }
 \affiliation{
Physical~Sciences~Department \\ Greenville~Technical~College \\ Greenville,
SC, 29606, USA}
\email{Wayne.Hayes@gvltec.edu}   

\author{J. R. Manson}
\affiliation{
Department of Physics and Astronomy, Clemson University \\ Clemson,
South Carolina 29634, U.S.A.}
\email{jmanson@clemson.edu}   

\date{\today}

\begin{abstract}
Recent experiments have  reported measurements of  rainbow scattering features in the angular distributions of hyperthermal Ar colliding with LiF(001)~[Kondo {\em et al.}, J. Chem. Phys. {\bf 122}, 244713 (2005)].
A semiclassical theory of atom-surface collisions recently developed by the authors that includes multiphonon  energy transfers  is used to explain the  temperature dependence of the measured scattered angular distribution spectra.
\end{abstract}
\maketitle


\section{Introduction} \label{intro}

Recently, Kondo {\em et al.} published an extensive series of experimental results on the scattering of well-defined incident beams of hyperthermal Ar atoms from clean LiF(001) surfaces.~\cite{Kondo-05,Kondo-06}  The data consist of angular distributions as a function of final scattering angle for an experimental apparatus having a fixed angle of $90^\circ$ between the incident beam and the detector position, and at each final angle all scattered particles were detected regardless of translational energy.  In contrast to many other measurements involving hyperthermal surface scattering of heavy rare gas atoms,~\cite{comment1} these measurements are noteworthy because they were carried out for a large range of incident energies, from 315 to 705 meV, and temperatures ranging from 300 to 573 K.  Under many conditions, and particularly for scattering in the plane of the $\langle 100 \rangle$ direction, the measured angular distributions exhibited two broad, but quite distinct peaks attributed to classical rainbow scattering caused by the corrugation of the LiF(001) surface.

In this paper we wish to analyze the Ar/LiF(001) data using using a new semiclassical atom-surface scattering theory that has been demonstrated to explain experimental measurements for hyperthermal energy scattering of rare gas atoms from liquid metal surfaces.~\cite{Hayes-12,Hayes-14}  This theory, briefly described in somewhat more detail in Sec.~\ref{theory} below, uses an interaction potential that is relatively straightforward and simple, a corrugated hard repulsive wall that allows for phonon vibrations due to the thermal motions of the underlying surface atoms.  Thus, although the interaction potential as a function of the $z$-coordinate normal to the surface is simple, the surface periodicity is retained and the the exchange of multiple phonons in the classical scattering limit is treated  realistically.
The results of our calculations explain the measured temperature dependence of the angular distributions, and show that at fixed incident energy, but with increasing surface temperature the increasing inelastic background broadens the rainbow peaks and makes them less distinct.

As a function of increasing incident energy it was noticed that the angular separation between the two rainbow peaks increased, giving the appearance of a surface corrugation height that decreased with energy.~\cite{Kondo-05}  However, it was predicted some years ago that such behavior of rainbow features could be the result of refraction of the incident and final scattering beams by the attractive adsorption potential well in front of the repulsive barrier.~\cite{Cole}

The interesting energy dependence of these data has been analyzed by Pollak {\em et al.} using a classical atom surface scattering perturbation theory that includes energy transfer to the phonons via a Langevin thermal bath and with the surface periodicity described by a one-dimensional corrugation function that contains multiple Fourier components.~\cite{Pollak-09,Pollak-12}
By including the Langevin description of energy transfer between projectile and surface, they showed that the Ar/LiF(001) angular distributions could be quite well replicated by a single, fixed, one-dimensional surface corrugation potential that included an attractive adsorption well and a very strong second order sinusoidal Fourier component.   Thus, they showed that the decrease with increasing incident energy in the angular separation of the two rainbow peaks could be explained by a single, fixed corrugation function, and did not  imply a corrugation that changed with incident energy.   However, their corrugation function contained not only a sinusoidal term describing one-dimensional periodicity with the LiF lattice constant, but also an even stronger second harmonic term with periodicity half that of the lattice constant.  The use of a corrugation function with strong higher-order Fourier components was originally proposed by Tully, with the explanation that the higher order terms mimicked the dynamical behavior of the interaction potential, i.e., they helped to account for the fact that the interaction potential can vary significantly depending on the height above the surface where the incoming projectile is deflected from the surface within the periodic unit cell.~\cite{Tully-90}  In a subsequent extension of this theory to a second order classical scattering perturbation treatment Pollak {\em et al.} have demonstrated that observed decrease in rainbow peak separation with increasing incident energy can be explained  with a one dimensional corrugation containing only a single Fourier component, but is dependent on the interaction potential having an appropriate attractive adsorption well.~\cite{Pollak-14}
A very general molecular dynamics approach which includes fully three-dimensional scattering has recently been reported.~\cite{Azuri}  Although at its present state of implementation calculations with this treatment shows only qualitative agreement with the Kondo data of Refs.~[\cite{Kondo-05}]~and [\cite{Kondo-06}],  they clearly demonstrate the shifts of the rainbow peaks with increasing incident energy.

In this paper we use a potential that is purely repulsive, without an attractive well, and although our calculations can match the energy dependence of the Ar/LiF(001) data, it requires a corrugation that decreases in height as the incident energy is increased.  In the light of Refs. [\cite{Cole,Pollak-09,Pollak-14,Azuri}] discussed above, this is clearly caused by the absence of an attractive well in the potential.  However, the temperature dependence of these data, which was not examined in Refs. [\cite{Pollak-09,Pollak-14}], is well explained by our theory of energy exchange with a repulsive barrier.  Also presented are several calculations for energy-resolved scattering spectra that illustrate how more detailed information on the interaction potential can be obtained from such measurements, as opposed to that provided by  measurements of the more restrictive angular distributions.

\section{Theory} \label{theory}

The theory used in this paper is fully developed in Ref.~[\cite{Hayes-14}], so we present only a few salient details here.   The starting point is the generalized Fermi golden rule for the transition rate of a projectile with initial momentum ${\bf p}_i$ going to momentum state ${\bf p}_f$ after the collision
\begin{eqnarray} \label{golden}
w({\bf p}_f,{\bf p}_i) ~=~ \frac{2 \pi}{\hbar}
\left\langle   \sum_{\{ n_f \}}   |T_{fi}|^2 \delta({\cal E}_f - {\cal E}_i)       \right \rangle
~,
\end {eqnarray}
where $ T_{fi}$ is the  matrix element of the transition operator  with respect to final and initial states of the  system of projectile plus target, $ {\cal E}_i$ and  $ {\cal E}_f$ are the initial and final  energies of the entire system, $\hbar$ is Planck's constant, the angular brackets signify an average over all initial states of the surface and the $ \sum_{\{ n_f \}} $ indicates a sum over all final states of the target.

The experimentally measured quantity is the differential reflection coefficient which gives the fraction of scattered intensity deflected into the small final energy interval $d E_f$ and small final solid angle $d \Omega_f$ and is related to the transition rate by
\begin{eqnarray} \label{DRC}
\frac{d R}{d E_f d \Omega_f}  ~\propto ~  \frac{p_f}{p_{iz}}~
w({\bf p}_f,{\bf p}_i)
~,
\end {eqnarray}
where on the right hand side the factor $p_f$ comes from the Jacobian relating phase space to energy space, and the perpendicular component of the incident momentum $p_{iz}$ is proportional to the incident particle flux.

Applying  the generalized Fermi golden rule of Eq.~(\ref{golden}) to scattering from a vibrating hard corrugated wall within the eikonal approximation of Levi {\em et al.}~\cite{Levi-75}, under conditions where large numbers of phonons are excited, leads to a transition rate expressed as a double integral over the surface of the target
\begin{eqnarray} \label{G3x}
w({\bf p}_f,{\bf p}_i) ~=~  \mathcal{L}~ \left(\frac{ p_{fz}}{m}\right)^2
\frac{2 \pi \hbar }{\sqrt{4 \pi\Delta E_0 k_B T_S}} ~~\exp{\left\{ -\frac{(E_f-E_i+\Delta E_0)^2}{4 \Delta E_0 k_B T_S}\right\}}
\\ \nonumber
\times ~~
 ~ \int ~d {\bf R} ~ \int ~d {\bf R^\prime} ~
e^{-i {\bf P} \cdot ({\bf R}-{\bf R^\prime})/ \hbar} ~
e^{-i \Delta p_z [\xi({\bf R}) -\xi({\bf R^\prime}) ]/ \hbar}
~\exp{\left\{-\frac{\Delta E_0  k_B T_S ( {\bf R} - {\bf R^\prime})^2 }{2 \hbar^2 v_R^2}\right\}}
~.
\end {eqnarray}
In Eq.~(\ref{G3x}) the three-dimensional recoil energy $\Delta E_0$ is related to the momentum vector ${\bf p}_f-{\bf p}_i$ through
$\Delta E_0 = ({\bf p}_f-{\bf p}_i)^2/2 M_C$ where $M_C$ is the effective mass of the target, ${\bf P}$ is the component of the momentum vector parallel to the surface, and $\xi({\bf R})$ is the surface corrugation function.  The mass of the projectile is $m$, $T_S$ is the surface temperature, $k_B$ is Boltzmann's constant, and $v_R$ is a velocity parameter~\cite{Brako,Brako-2} arising from the law of conservation of parallel momentum  for all phonons transferred.
The proportionality constant $\mathcal{L}$ involves only factors of the quantum confinement length.

The effective  mass $M_C$ is a measure of how many LiF molecules are involved in the recoil of the surface as a result of the collision with the Ar projectile.
The parameter $v_R$ is a weighted average over phonon speeds parallel to the surface~\cite{Brako,Brako-2} and its significance, as well as suggestions of how to best measure it, are discussed in Refs.~[\cite{Hayes-07,Hayes-07-2,Hayes-11-2}].

Eq.~(\ref{G3x}) is a semiclassical approximation, but evaluated in the limit in which large numbers of phonons are transferred in the collision, so many phonons that the specific details of the phonon distribution function become unimportant and the energy transfer is expressed through the recoil function $\Delta E_0$.  For large projectile energies the coherence length for quantum interference becomes negligibly small, and the double integral over the surface can be replaced by integrals over a single unit cell, multiplied by the number of unit cells which is simply a large constant.

An interesting limiting case arises upon assuming that the surface is flat, which means the corrugation function  $\xi({\bf R})$ is constant.   Taking the classical limit in this case
the integrals can be carried out leading to
\begin{eqnarray} \label{W4}
w_{}({\bf p}_f,{\bf p}_i) ~ \propto ~
 ~ \frac{1}{ (4 \pi k_B T_S \Delta E_0)^{3/2}} ~ p_{fz}^2 ~
\exp\left\{ - \frac{\left(E_f - E_i  + \Delta E_0  \right)^2 + 2  v_R^2 {\bf P}^2}
{4 k_B T_S \Delta E_0} \right\}
~,
\end {eqnarray}
With the exception of the factor of $p_{fz}^2$ Eq.~(\ref{W4}) is the form first derived by Brako and Newns.~\cite{Brako,Brako-2,Levine,Manson-91,Himes}
The original Brako-Newns theory neglected the effects of the elastic, or non-vibrating, part of the interaction potential.  Later work has shown that the elastic part of the potential introduces a multiplicative factor which is a squared matrix element of the elastic potential.  The factor of $p_{fz}^2$  comes from the this matrix element evaluated for a hard repulsive wall as introduced by the eikonal approximation.
Eq.~(\ref{W4}) has been successfully used to describe the energy and temperature dependence of atom-surface scattering for many systems observed under classical conditions.~\cite{Hayes-07,Hayes-07-2,Hayes-11-2,Celli-91}  However, because it describes the surface as a flat repulsive wall with no static corrugation, it is incapable of explaining rainbow features.  The theory embodied in Eq.~(\ref{G3x}) is a generalization of the Brako-Newns expression (\ref{W4}) to include surface corrugation.

\section{Comparison with Experiments} \label{comparisons}

For the calculations reported here we use a one dimensional model for the surface corrugation with a sinusoidal corrugation function
\begin{eqnarray} \label{corr}
\xi({\bf R}) ~\longrightarrow ~ \xi(x) ~ = ~ h b \cos\left(\frac{2 \pi x}{b}\right)
~.
\end {eqnarray}
The lattice parameter of LiF is $a=4.02$~\AA~\cite{Kondo-05,Ekinci}, thus calculations based on Eq.~(\ref{G3x}) depend on three parameters.  These are the dimensionless corrugation parameter $h$, the effective mass  $M_C$ of the LiF, and the velocity parameter $v_R$.
It is also noted that even though the corrugation is one-dimensional the scattering theory of Eq.~(\ref{G3x}) is fully three-dimensional and energy exchange involves all three directions on the surface.

\begin{figure}
\includegraphics[width=5.5in]{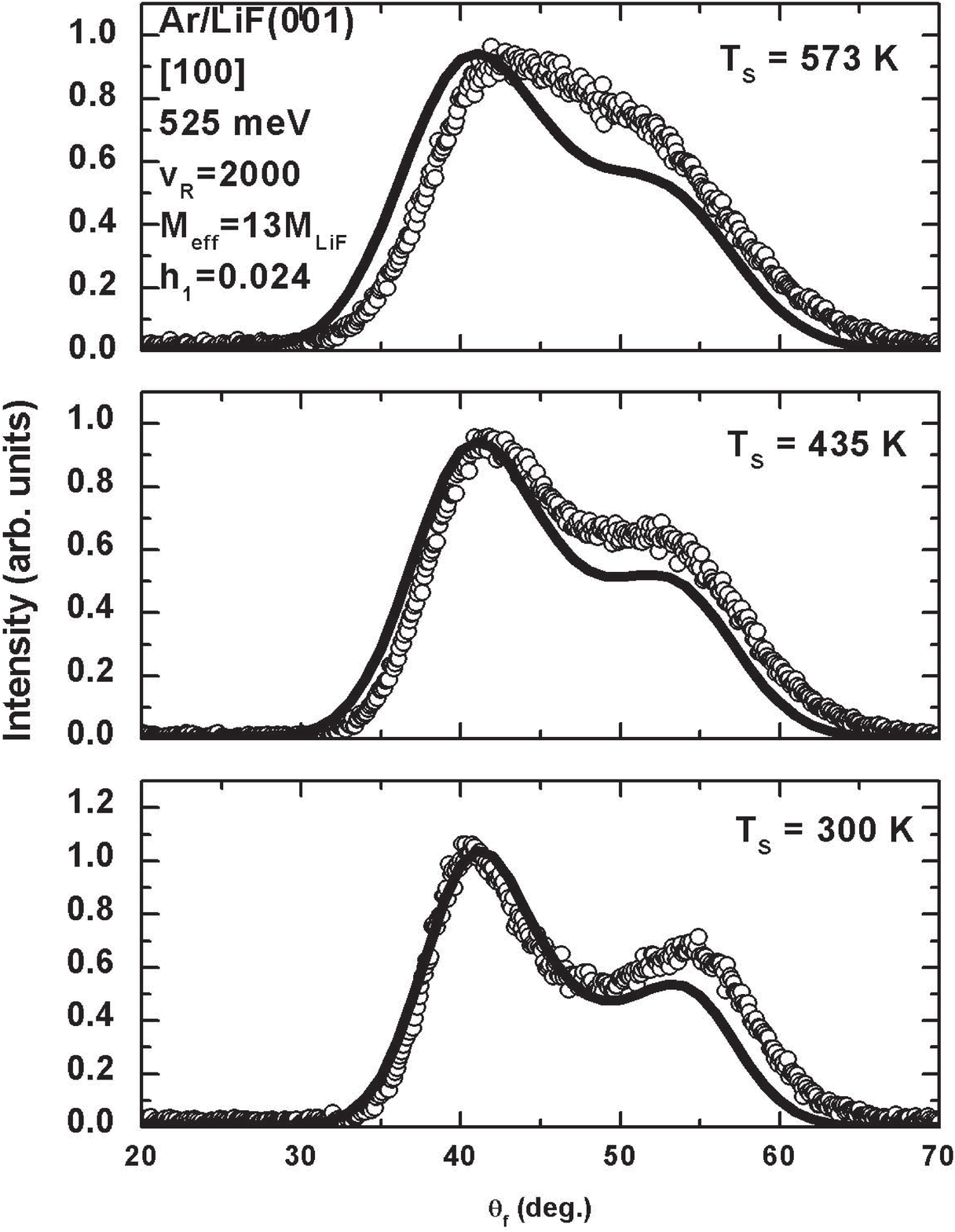}
\caption{Temperature dependent angular distribution spectra for Ar scattering from LiF(001)$\langle 100 \rangle$ with an incident energy of 525 meV and three different surface temperatures as marked.  Solid curves are the present calculations with a corrugation parameter $h=0.024$ and the experimental points are from Ref.~[\cite{Kondo-05}].
}
\label{T-dep100}
\end{figure}

Angular distribution spectra  for in-plane scattering of Ar from LiF(001) along the $\langle 100 \rangle$ direction are compared with calculations as a function of final detector angle $\theta_f$ in Fig.~\ref{T-dep100}.  The spectra were measured at three different surface temperatures of 300, 435 and 573 K as marked and the incident energy is $E_i = 525$ meV.  At the lowest temperature of 300 K there are two clearly distinct rainbow peaks, asymmetrically positioned about the specular position $\theta_f = \pi/4$, and those two peaks differ significantly in intensity.  As the temperature is raised the increased inelastic scattering causes the two-peak feature to become less distinct and the whole scattered distribution broadens slightly.

It is of interest to briefly discuss the expected rainbow pattern from an elastic, mirror-reflecting surface.
For scattering in the plane of incidence with a symmetrically shaped corrugation,  such as the cosine function of Eq.~(\ref{corr}), the rainbow positions would be symmetrically positioned about the specular position.   If the angle of incidence $\theta_i$ is fixed, the rainbow angles corresponding to mirror reflection from the inflection regions of the cosine corrugation of  Eq.~(\ref{corr}) would appear at the two final angles $\theta_f = \theta_i \pm 2 \tan^{-1} (2 \pi h)$.  For the the fixed source-detector geometry in the experiments considered here the mirror reflection rainbow angles of the cosine corrugation would appear symmetrically about the specular position with $\theta_f = \pi/4 \pm \tan^{-1} (2 \pi h)$.  Clearly the data shown in Fig.~\ref{T-dep100} do not exhibit the symmetry of a mirror surface.  The subspecular rainbow peak appears much closer to the specular position $\theta_f=\pi/4$ than the supraspecular peak.  The large inelastic energy exchange between the Ar and the LiF surface causes the rainbow pattern to appear quite differently from the form predicted by mirror reflection.  The softness present in an interaction potential more realistic than the present hard wall repulsion has also been demonstrated to contribute to the shifts of the rainbow peaks,~\cite{Pollak-09,Pollak-12,Pollak-14,Azuri} but in the present calculation all of the shift is caused by inelastic energy transfer.

The calculations from Eq.~(\ref{G3x}) are shown as solid curves in Fig.~\ref{T-dep100}.  These are obtained from the differential reflection coefficient of Eq.~(\ref{DRC}) after integrating over all final energies.  In the integral over final energies a factor of $1/\sqrt{E_f}$ is introduced to account for the $1/v_f$ velocity dependence of the experiment which uses a density detector.  The calculations use the following parameter values: $h = 0.024$, $M_C$ is 13 times the 25.9 amu mass of LiF or 337 amu,  and  $v_R=2000$ m/s.  These same mass and $v_R$ values are used for all calculations shown in this paper.  The calculated intensities are not strongly affected by the choices of these latter two parameters.  If $M_C$ is changed by 20\% and/or $v_R$ is changed by 50\% the results differ very little, hence the results are not critically dependent on the fitting of these two parameters.
The repeat distance for the one-dimensional corrugation was chosen as  $b = a/2 = 2.01$~\AA.  This choice is motivated by the fact that for the LiF(001) surface the predominant feature causing the corrugation is the bumps presented by the large F ions, which form a square array, with sides of length $a/\sqrt{2} = 2.84$~\AA, rotated $45^\circ$ with respect to the LiF(001) unit cell.
He atom diffraction peaks in the $\langle 100 \rangle$ direction are separated by parallel wave vectors of 3.1~\AA$^{-1}$ which corresponds to diffraction from rows with a spacing of $b=a/2$.~\cite{Kondo-05,Ekinci,Boato-76}
Nevertheless, this choice is not important in the calculations presented here for the following reason: as mentioned above for the case of classical scattering from a hard mirror surface the positions of the rainbow peaks are defined by the slope of the corrugation at its inflection point, and from Eq.~(\ref{corr}) this is $d \xi /dx = \pm 2 \pi h$ and is independent of the repeat distance $b$, but depends on the dimensionless corrugation parameter $h$.  Because the present calculations are in the classical limit where quantum interference is negligible our calculations are similarly sensitive primarily to the corrugation parameter $h$.  Thus calculations using a repeat distance $b=a$ are nearly identical to those for $b=a/2$.

The calculations are  strongly sensitive to the choice of the dimensionless corrugation parameter, as will become evident in the discussion of below.  The value of the corrugation parameter $h= 0.024$ is chosen to give the best fit with the data at the lowest temperature of 300 K.
Because the data were reported in arbitrary units, the calculations are normalized to the data at a single point in each panel of Fig.~\ref{T-dep100}.
At all temperatures the calculations match the data well and appear to explain the smoothing and broadening of the rainbow features as the temperature increases.  The asymmetry in both angular position and relative intensity of the two rainbow peaks is well predicted, and the temperature broadening is also reasonably explained by the inelastic transfers built into this theory.

\begin{figure}
\includegraphics[width=5.5in]{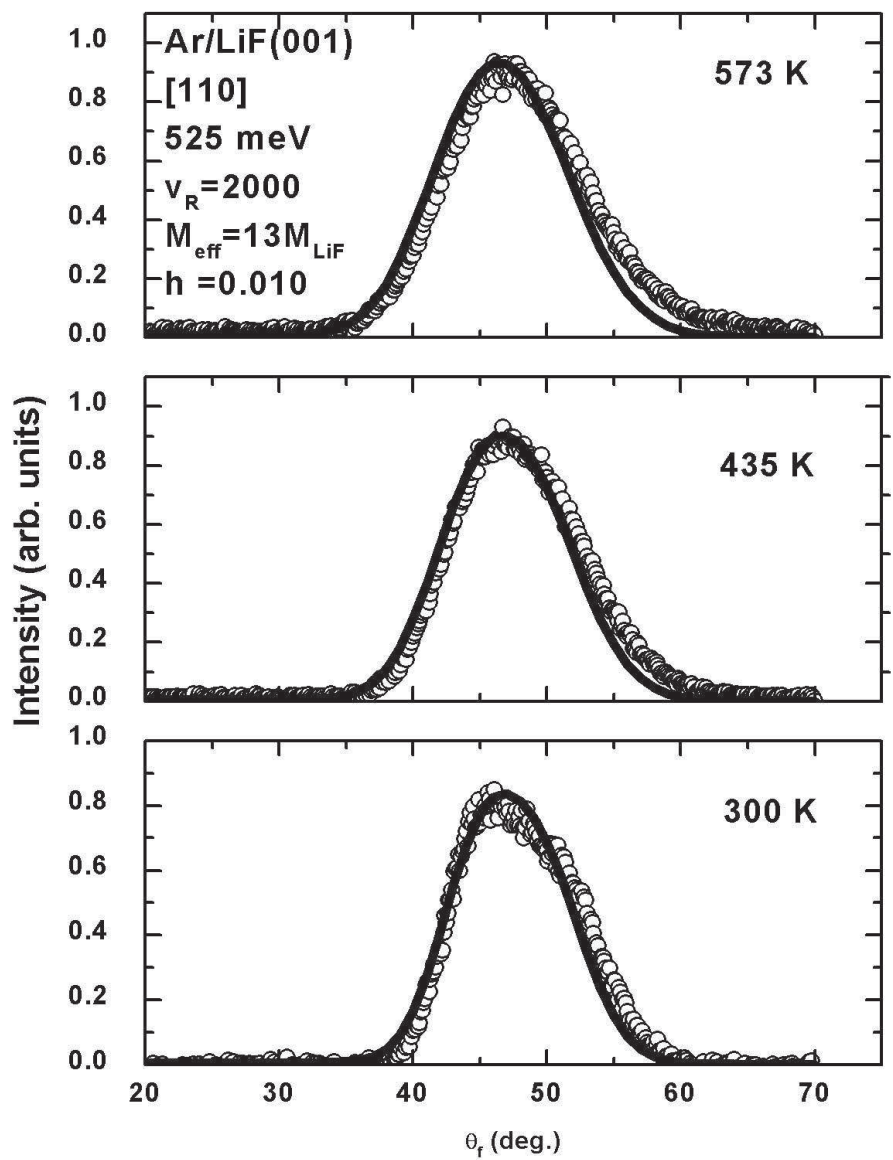}
\caption{Temperature dependent angular distribution spectra for Ar scattering from LiF(001)$\langle 110 \rangle$ with an incident energy of 525 meV and $h=0.01$.  All other conditions are the same as in Fig.~\ref{T-dep100}
}
\label{T-dep110}
\end{figure}

Fig.~\ref{T-dep110} shows the temperature dependence of the angular distribution spectra measured in the $\langle 110 \rangle$ direction for $E_i = 525$ meV.  The $\langle 110 \rangle$ direction lies along the squares formed by the fluorine atoms on LiF(001) so the one dimensional sinusoidal period is chosen to be $b = a/\sqrt{2} = 2.84$~\AA.  The three surface temperatures are the same as in Fig.~\ref{T-dep100}.   In this direction the expected two peaks of the rainbow pattern have merged together and appear as a single broad feature, an indication that the apparent corrugation strength is weaker.  The  calculations are similar to those of Fig.~\ref{T-dep100} except now the corrugation parameter takes the smaller value $h=0.01$.  With this corrugation the  temperature dependence of the data is explained rather well which indicates that the merging of the two expected rainbow peaks and the observed broadening of the single merged feature may be explained as a consequence of a rather large multiphonon energy transfer from the Ar to the surface.  This statement should be tempered with caution, however, because it has been shown that softness in the repulsion for a more realistic interaction potential, as well as the contribution of the attractive potential well, can also contribute to the shifts and broadening of these features.~\cite{Pollak-09,Pollak-12,Pollak-14}  The calculations also indicate that the apparent corrugation in the $\langle 110 \rangle$ direction is significantly smaller than in the $\langle 100 \rangle$ direction.

Although it is of interest to examine the temperature dependence of angular distributions, the following figures present calculations that show how substantially more detailed information about the collision process can be obtained from energy-resolved measurements.  Fig.~\ref{TE-dep100} shows the calculated temperature dependence of the scattered intensity as a function of final energy for the same conditions as for the angular distributions in Fig.~\ref{T-dep100} except that the detector position is held at the specular position, i.e., $\theta_i=\theta_f = 45^\circ$.
At all detector angles $\theta_f$ for which the scattered intensity is appreciable the calculated energy-resolved plots consist of a single, broad peak.
The position of the most probable final energy (the peak position) remains the same at all temperatures, but the most probable peak intensity (the peak height) decreases strongly with increasing temperature while the full width at half max (FWHM) increases.  This is the expected behavior since the calculations are unitary, implying that the scattered flux must equal the incoming flux.  Thus, as the temperature increases and distributes the scattered particles over a larger range of final energies the maximum scattered intensity must decrease in order to preserve the total flux.  It is also noticed that at higher temperatures the scattered intensity develops an increasingly asymmetric tail on the high energy side.  This indicates that at higher temperatures increasingly larger numbers of atoms are ``heated up" and some are predicted to leave the surface with substantially larger energies than the incident beam, although the stable peak position indicates that the most probable energy loss under these conditions remains at about 10\% of the incident energy.  The FWHM increases with temperature at a rate much faster than a proportionality to $\sqrt{T_S}$ that would be expected from examination of the smooth surface model of Eq.~(\ref{W4})  and this is apparently due to the rapidly increasing asymmetry in the high energy tail

It is of interest to note the sharp decrease in most probable intensity as a function of temperature in the energy-resolved spectra shown in Fig.~\ref{TE-dep100}.  The calculated angular distributions, such as shown in Figs.~\ref{T-dep100} and~\ref{T-dep110}, also show a decrease in maximum peak intensities, but not nearly so strong as for  typical energy-resolved spectra.  This is because an angular distribution  is a measurement that, at a given detector position, sums the intensities over all final energies, and again because of the unitarity condition this means that the temperature dependence will be less pronounced.

It has been demonstrated that the temperature dependence of an energy-resolved measurement can be directly related to the surface corrugation amplitude,~\cite{Hayes-12,Hayes-14,PollakManson} and this is illustrated in Fig.~\ref{TMP-dep100} which shows as open circles the most probable intensity of Fig.~\ref{TE-dep100} as a function of $T_S$.   Also shown in Fig.~\ref{TMP-dep100} as a solid curve is the expected behavior of approximately $1/T_S^{3/2}$ for the smooth surface model as seen from the prefactor to the exponential of Eq.~(\ref{W4}), and the dashed curve shows the expected behavior of a highly corrugated surface of discrete scattering centers which goes approximately as $1/T_S^{1/2}$.  Calculations for a surface with corrugation intermediate between these two extremes should give a locus of points on a curve lying between the $1/T_S^{3/2}$ and $1/T_S^{1/2}$ behaviors, and this is demonstrated  by the open circle points taken from Fig.~\ref{TE-dep100}.  The relative  position of the curve traced by those points is defined by the value of $h$ for the corrugation amplitude of the corrugation in Eq.~(\ref{corr}) above.  The proximity of the present calculations to the $1/T_S^{1/2}$ behavior indicates that this system, as expected, is rather highly corrugated.

As another example demonstrating the potential sensitivity of energy-resolved spectra to the parameters of the interaction potential, Fig.~(\ref{M-dep100}) shows calculations under the same conditions as Fig.~\ref{TE-dep100} at the temperature of 573 K, but with three different effective masses.  As stated above, the angular distributions of Figs.~\ref{T-dep100} and~\ref{T-dep110} are rather insensitive to the value chosen for the effective mass, and the calculated angular distributions under those conditions would be essentially unchanged if the mass were increased or decreased by as much as 20\%.  However, the situation is quite different for the energy resolved spectra, as  seen in Fig.~(\ref{M-dep100}) which shows, in addition to the spectrum for the $M_C$ value of 13 LiF masses, curves for $M_C=11$ and 15.  Increasing or decreasing $M_C$ by $\approx$20\% gives rise to most probable final energies that differ by as much as 50 meV which is a very readily measurable amount.  Thus it is very apparent that energy-resolved spectra are a much more sensitive way to sample the effective mass of the target.

\section{Discussion and Conclusions} \label{conclusions}

In this paper we examine a set of data for  angular distributions of hyperthermal energy Ar atoms scattering from the LiF(001) surface.  These data are analyzed with a new theory of atom scattering from corrugated surfaces under conditions for which the collision is rapid and results in the transfer of large amounts of energy via multiphonon excitations.  Previously, we showed that relatively straightforward temperature-dependent measurements on the maximum intensities of energy-resolved scattering spectra, taken under conditions of fixed incident and detector angles, could be used to estimate the corrugation height of the surface.~\cite{Hayes-12,Hayes-14}  The presently analyzed measurements are angular distributions, meaning that at each  angle all particles are detected regardless of final energy.  This provides less detailed information about the scattering process because of the integration over all final energies, but the detection process is simpler and can be followed over a range of final angles and thus is capable of measuring effects such as the classical rainbow peaks discussed here.

The angular distribution data exhibit a number of characteristic effects.  Many of the angular distributions show two distinct rainbow peaks, the signature of a simply corrugated surface with two inflection regions. However, these two peaks are not symmetrically positioned in angular separation with respect to the specular direction as would be expected for reflection from a mirror surface.  As the surface temperature is increased the rainbow pattern becomes less distinct and at the highest temperatures, and especially for the $\langle 110 \rangle$ direction, they merge into a single broad structure.

Within the confines of our simple interaction potential, the calculations indicate that the positional asymmetry of the two rainbow peaks, in both the high-symmetry $\langle 100 \rangle$ and $\langle 110 \rangle$ directions, can be largely accounted for due the inelastic transfers resulting from the exchange of energy.  This is consistent with previous work of others,~\cite{Pollak-09,Pollak-14} but there they found that some of the asymmetry is due to softness and the attractive potential well that would appear in a more realistic interaction potential.  Because our interaction potential lacks softness and has no well, it is possible that this could result in over- or underestimation of our chosen interaction parameters.

A significant contribution of the present work is that it explains the temperature dependence of the measured angular distributions.  In both symmetry directions, our calculations match the spectra reasonably well at low temperatures.  As the temperature is increased our calculations show how the rainbow peaks become less distinct and gradually merge into a single, broad feature as a consequence of larger energy transfer with the surface.

It should be noted that this work is based on using a one-dimensional corrugation, although the scattering theory and calculations are fully three-dimensional.  The real corrugation of the LiF(001) surface is two-dimensional, which could have distinct and qualitatively different scattering effects as has been indicated by recent work.~\cite{Azuri}
Although this work, as well as previous analysis,~\cite{Pollak-09,Pollak-14} arrives at  physically reasonable explanations of the  Ar/LiF(001) angular distributions, these results need to be verified with a more realistic three-dimensional treatment.  Work is presently in progress to carry out such three-dimensional calculations within the framework of the present theory.

\begin{figure}
\includegraphics[width=5.5in]{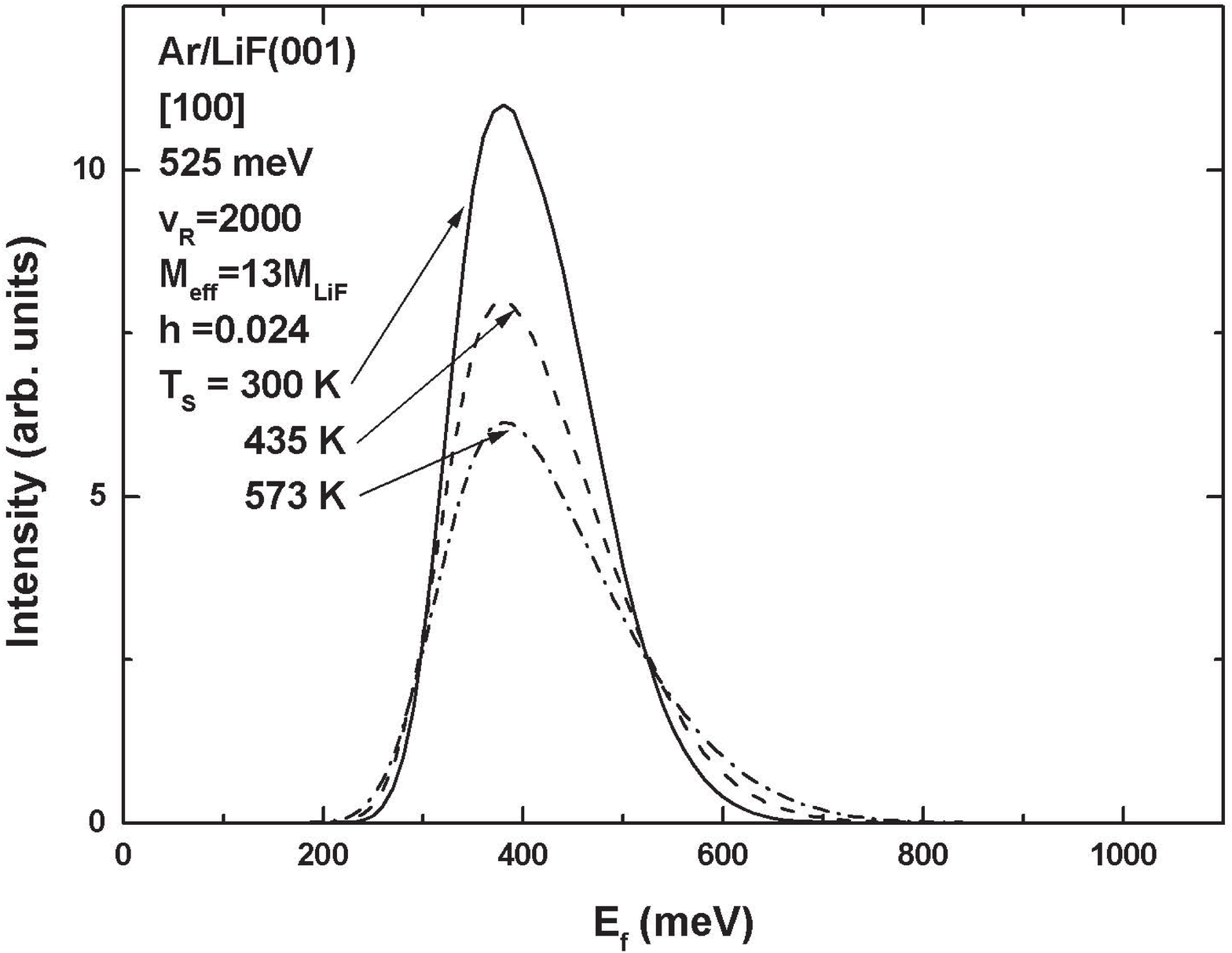}
\caption{Calculations showing temperature dependence of  energy-resolved spectra for Ar scattering from LiF(001)$\langle 100 \rangle$ at an incident energy of 525 meV.  The incident and final angles are fixed at $\theta_i = \theta_f= 45^\circ$, while all other conditions are the same as for Fig.~\ref{T-dep100}.
}
\label{TE-dep100}
\end{figure}

\begin{figure}
\includegraphics[width=5.5in]{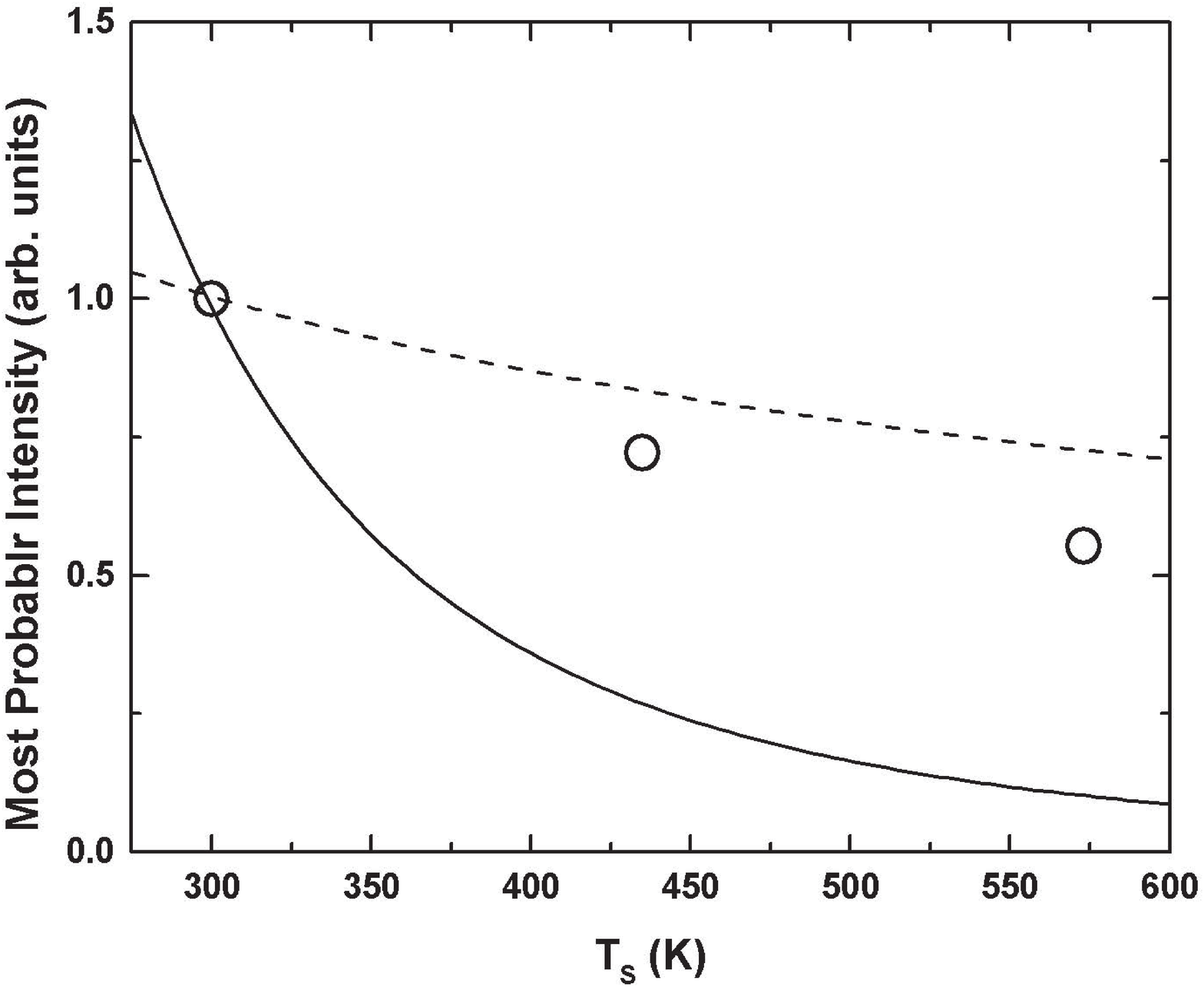}
\caption{Temperature dependence of the most probable intensity of the energy-resolved spectra
shown in Fig.~\ref{TE-dep100}, compared to that for smooth and discrete surfaces, shown as solid and dashed curves, respectively.
}
\label{TMP-dep100}
\end{figure}

\begin{figure}
\includegraphics[width=5.5in]{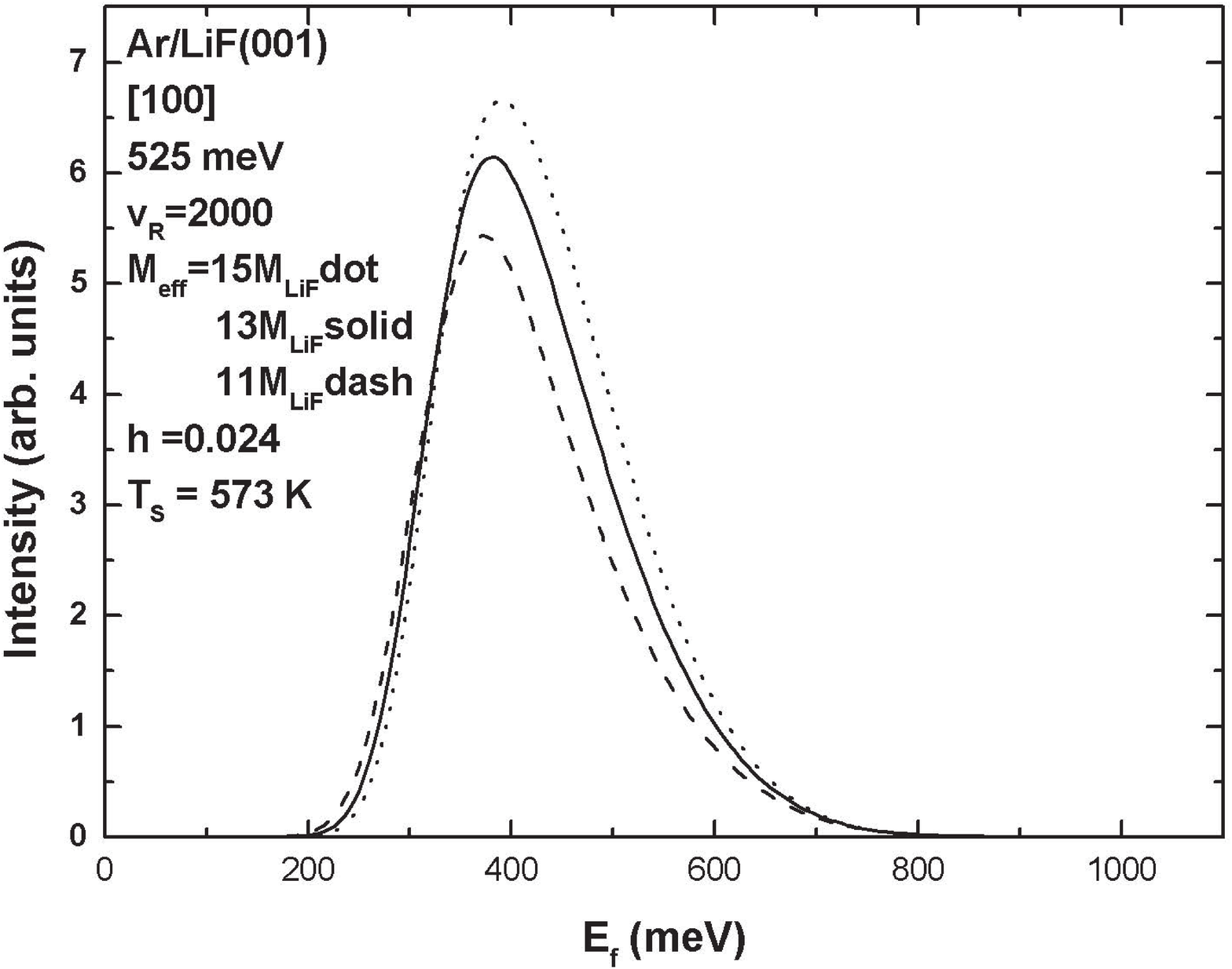}
\caption{Calculations of energy-resolved spectra for Ar scattering from LiF(001)$\langle 100 \rangle$
at $T_S = 573$ K showing dependence on effective surface mass $M_C$.  Solid curve: same as Fig.~\ref{TE-dep100}; dashed curve: with effective mass reduced by the mass of 2 LiF molecules; dotted curve: with mass increased by two LiF molecules.
}
\label{M-dep100}
\end{figure}

In addition to direct comparison with the experimentally measured angular distributions, we showed in Figs.~\ref{TE-dep100}-\ref{M-dep100} several calculations of predicted behavior of energy-resolved spectra taken at fixed incident and detector angles.  These examples demonstrate that the energy-resolved spectra are substantially more sensitive to experimental parameters than are the energy-integrated angular distributions. This indicates that the best types of experiments to give the most information on the parameters characterizing the interaction are measurements of energy-resolved spectra at each combination of final and initial angles.
\\~\\
{\bf Acknowledgments}

The authors would like to thank Prof. Takahiro Kondo for helpful advice, and for providing files of the experimental data shown here.  One of us (JRM) would like to thank Eli Pollak and Salvador Miret-Art{\'e}s for helpful discussions.

\newpage

\newpage
\listoffigures


\newpage

\begin{thebibliography}{199}

\bibitem{Kondo-05}  Takahiro Kondo, Hiroyuki S. Kato, Taro Yamada, Shigehiko Yamamoto and Maki Kawai, {\em Rainbow scattering of CO and N$_2$ from LiF(001)}, J. Chem. Phys. {\bf 122}, 244713 (2005).

\bibitem{Kondo-06}  T. Kondo, H. S. Kato, T. Yamada, S. Yamamoto and M. Kawai,
 {\em Effect of the molecular structure on the gas-surface scattering studied by supersonic molecular beam},
 Eur. Phys. J. D {\bf 38}, 129 (2006).

\bibitem{comment1}  For an extensive reference list of the literature in hyperthermal atom-surface scattering experiments, see Refs.~[\cite{Pollak-09}] and~[\cite{Pollak-12}].


\bibitem{Hayes-12}  W. W. Hayes and J. R. Manson,
{\em Determination of the Surface Corrugation Amplitude from Classical Atom Scattering },
 Phys. Rev. Lett. {\bf 109}, 063203 (2012).

\bibitem{Hayes-14}  W. W. Hayes and J. R. Manson,
{\em Classical and semiclassical theories of atom scattering from corrugated surfaces },
Phys. Rev. B {\bf 89 }, 045406 (2014).

\bibitem{Cole}  James R. Klein and Milton W. Cole,
{\em On the energy dependence of rainbow peaks in classical atom/surface scattering},
 Surface Science Letters {\bf 81}, L319 (1979).

\bibitem{Pollak-09} Eli Pollak, Jeremy M. Moix and Salvador Miret-Art{\'e}s,
 {\em Classical theory for asymmetric in-plane atom surface scattering},
 Phys. Rev. B {\bf 80}, 165420 (2009).


\bibitem{Pollak-12}  S. Miret-Art{\'e}s and E. Pollak,
{\em Classical theory of atom-surface scattering: The rainbow effect},
Surf. Sci. Reports {\bf 67}, 161 (2012).


\bibitem{Tully-90}  J. C. Tully,
{\em Washboard model of gas surface scattering},
J. Chem. Phys. {\bf 92}, 680 (1990).

\bibitem{Pollak-14}  Yun Zhou, Eli Pollak and Salvador Miret-Art{\'e}s
 {\em Second order classical perturbation theory for atom surface scattering: Analysis of asymmetry in the angular distribution},
 J. Chem. Phys. {\bf 140}, 024709 (2014).

\bibitem{Azuri} Asaf Azuri and Eli Pollak, {\em On the fly first principles study of the classical scattering of an Ar atom from the LiF(100) surface}, J. Chem. Phys. {\bf 139}, 044707 (2013).

\bibitem{Levi-75}  U. Garibaldi, A. C. Levi, R. Spadacini and G. Tommei,
{\em Quantum-theory of atom-surface scattering-diffraction and rainbow},
Surface Science {\bf 48}, 649 (1975).

\bibitem{Brako}  R. Brako and D. M. Newns,
{\em Differential crosss-section for atoms inelastically scattered from surfaces},
 Phys. Rev. Lett. {\bf 48}, 1859 (1982).

\bibitem{Brako-2} R. Brako and D. M. Newns,
{\em Energy and angular-distribution of atoms scattered from surfaces},
Surf. Sci. {\bf 117}, 42 (1982).


\bibitem{Hayes-07}  W. W. Hayes and J. R. Manson,
{\em Argon scattering from Ru(0001): Calculations and comparison with experiment },
Phys. Rev. B {\bf 75}, 113408 (2007).

\bibitem{Hayes-07-2}  W. W. Hayes and J. R. Manson,
{\em Rare gas collisions with molten metal surfaces },
J. Chem. Phys. {\bf 127}, 164714 (2007).

\bibitem{Hayes-11-2}  W. W. Hayes and J. R. Manson,
{\em Theoretical determination of the effective velocity parameter in atomic and molecular scattering from surfaces},
Phys. Rev. B {\bf 74}, 073413 (2006).



\bibitem{Levine}  H.-D. Meyer and R. D. Levine,
{\em Multiphonon energy-transfer in atom surface scattering},
Chem. Phys. {\bf 85}, 189 (1984).

\bibitem{Manson-91}  J. R. Manson,
{\em Inelastic-scattering from surfaces},
Phys. Rev. B {\bf 43}, 6924 (1991).

\bibitem{Himes}  J. R. Manson, V. Celli and D. Himes,
{\em Multiphonon scattering from surfaces},
 Phys. Rev. B {\bf 49}, 2782 (1994).

\bibitem{Celli-91}  V. Celli, D. Himes, P. Tran, J. P. Toennies, C. W\"oll, G. Zhang,
 {\em Multiphonon processes in atom-surface scattering},
 Phys. Rev. Lett. {\bf 66}, 3160 (1991).

\bibitem{Ekinci}  Y. Ekinci and J. P. Toennies, {\em Thermal expansion of the LiF(001) surface}, Surf. Sci. {\bf 563}, 127 (2004).

\bibitem{Boato-76}  G. Boato, P. Cantini and L. Mattera, {\em A Study of the (001)LiF Surface at 80 K by Means of Diffractive Scattering of He and Ne Atoms at Thermal Energies}, Surf. Sci. {\bf 55}, 141 (1976).

\bibitem{PollakManson}  Eli Pollak and J R Manson, {\em Temperature dependence in atom-surface
scattering}, J. Phys.: Condens. Matter {\bf 24}, 104001 (2012).



\end{thebibliography}
\end{document}